\documentclass[useAMS,usenatbib]{mn2e}

\usepackage{graphicx}
\usepackage{amssymb}
\usepackage{longtable}
\usepackage{natbib}
\usepackage{rotating}
\usepackage{color, colortbl}
\definecolor{Gray}{gray}{0.9}
\bibpunct{(}{)}{;}{a}{}{,} 
\title[Studying the evolution of galaxies in compact groups. ]{Studying the evolution of galaxies in compact groups over the past 3 Gyr -- II. The importance of environment in the suppression of star formation}
\author[T. Bitsakis et al.]{T. Bitsakis$^{1}$\thanks{E-mail: tbitsakis@astro.unam.mx}, D. Dultzin$^{1}$, L. Ciesla$^{2,3}$, T. D\'iaz-Santos$^{4}$, P. Appleton$^{5}$, \newauthor V. Charmandaris$^{3,6,7}$, Y. Krongold$^{1}$, P. Guillard$^{8,9}$, K. Alatalo$^{5,10}$\thanks{Hubble fellow}, A. Zezas$^{3,11,12}$, \newauthor J. Gonz\'alez$^{1}$, and L. Lanz$^{13}$ \\
$^{1}$Instituto de Astronom\'ia, Universidad Nacional Aut\'onoma de M\'exico, P.O. 70-264, 04510 D.F., Mexico\\
$^{2}$Laboratoire AIM, CEA/DSM/IRFU, CNRS, Universit\'e Paris-Diderot, 91190 Gif, France\\
$^{3}$Department of Physics, University of Crete, Heraklion 71003, Greece\\
$^{4}$N\'ucleo de Astronom\'ia de la Facultad de Ingenier\'ia, Universidad Diego Portales, Av. Ej\'ercito Libertador 441, Santiago, Chile\\
$^{5}$NASA Herschel Science Center, 770 S. Wilson Ave., Pasadena, CA 91125, USA\\
$^{6}$Institute for Astronomy, Astrophysics, Space Applications \& Remote Sensing, National Observatory of Athens, GR-15236, Penteli, Greece\\
$^{7}$Chercheur Associ\'e, Observatoire de Paris, F-75014, Paris, France\\
$^{8}$Institut d'Astrophysique de Paris, CNRS, UMR 7095, 98 bis Boulevard Arago, 75014 Paris, France\\
$^{9}$Sorbonne Universit\'es, UPMC Universit\'e Paris 06, 4 place Jussieu, 75005 Paris, France\\
$^{10}$Observatories of the Carnegie Institution of Washington, 813 Santa Barbara Street, Pasadena, CA 91101, USA\\
$^{11}$Harvard-Smithsonian Center for Astrophysics, Cambridge, MA 02138, USA\\
$^{12}$Foundation for Research and Technology - Hellas (FORTH), Heraklion 71003, Greece\\
$^{13}$Infrared Processing and Analysis Center, California Institute of Technology, MC100-22, Pasadena, California 91125, USA
}
\begin{document}


\pagerange{\pageref{firstpage}--\pageref{lastpage}} \pubyear{2015}

\maketitle

\label{firstpage}

\begin{abstract}
We present an in depth study on the evolution of galaxy properties in compact groups over the past 3 Gyr. We are using the largest multi-wavelength sample to-date, comprised  1770 groups (containing 7417 galaxies), in the redshift range of 0.01$<$z$<$0.23. To derive the physical properties of the galaxies we rely on ultraviolet (UV)-to-infrared spectral energy distribution modeling, using {\sevensize CIGALE}. Our results suggest that during the 3 Gyr period covered by our sample, the star formation activity of  galaxies in our groups has been substantially reduced (3-10 times). Moreover, their star formation histories as well as their UV-optical and mid-infrared colors are significantly different from those of field and cluster galaxies, indicating that compact group galaxies spend more time transitioning through the green valley. The morphological transformation from late-type spirals into early-type galaxies occurs in the mid-infrared transition zone rather than in the UV-optical green valley. We find evidence of shocks in the emission line ratios and gas velocity dispersions of the late-type galaxies located below the star forming main sequence. Our results suggest that in addition to gas stripping, turbulence and shocks might play an important role in suppressing the star formation in compact group galaxies. 
\end{abstract}


\section{Introduction}
Over the past three decades, numerous studies indicated that interactions between galaxies can affect their evolution in terms of morphology, gas and dust content, as well as star formation and nuclear activities. While it has been shown that interactions can enhance star formation by fueling gas towards the central regions of galaxies and/or increase their gas content via accretion and merging \citep[e.g.][and references therein]{Struck99}, a number of studies indicated that the picture is more complex and other physical phenomena, such as shocks and turbulence also play a role \citep[e.g.][]{Appleton06, Boselli08, Appleton13, Alatalo14}. Since these processes can influence the evolution of galaxies, the environment of galaxies is very relevant. Compact groups of galaxies have been considered ideal targets to study environmental effects. Due to their high galaxy densities and low velocity dispersions, their members have experienced numerous and frequent interactions during their lifetimes \citep{Hickson92}. Some studies have shown that a large fraction of cluster galaxies might have been pre-processed in groups \citep[e.g.][]{Cortese06,Eckert14}, thus stressing even more the importance of  understanding the mechanisms driving the evolution of galaxies in different kinds of groups. 

The most studied compact group sample was the one compiled by \citet{Hickson82}, which consists of 100 local (z$<$0.06) compact groups (hereafter HCGs), containing 451 galaxies. Further analysis using spectroscopic information reduced this sample to 96 groups \citep{Hickson92}. Detailed studies revealed that compact groups contain about half of the late-type galaxy fraction that is observed in the field and almost double of what is seen in clusters \citep{Hickson82}, and that they display ubiquitous signs of tidal interactions \citep{Mendes94}. \citet{Verdes01} reported gas deficiencies in the individual galaxies, while \citet{Borthakur10} showed that the most evolved groups contain a diffuse component of HI gas in their intra-group medium. They proposed that as the groups evolve, neutral gas is ejected by some mechanism into the group environment.

More recently, the advent of high-resolution infrared telescopes (such as the {\it Spitzer}, {\it Herschel}, and the Wide-field Infrared Survey Explorer; {\it WISE}) made it possible to study  the dust obscured star formation activity, as well as the cold dust properties of individual group galaxies. \citet{Johnson07} and \citet{Walker12} examined a sample of 12 HCGs and proposed that the observed deficit of galaxies in a given range of mid-IR colours could have been due to their rapid evolution from the star forming to the quiescent sequences. Warm H$_{2}$ and [C{\sevensize II}] imaging revealed that the well known ``Stephan's Quintet'' system contains intergalactic material shock-heated by collisions within the group \citep{Appleton06, Cluver10, Appleton13}. In a more comprehensive study of two dozen HCGs, \citet{Cluver13} found over 20 per cent of the groups contained galaxies with enhanced shocked H$_{2}$ emission, and these fell primarily in mid-infrared green valley, a color space defined by {\it Spitzer} IRAC colors. More recent {\it Herschel} and {\it CARMA} observations from \citet{Alatalo15b} have shown that many of these same galaxies contain molecular gas that is unable to form stars efficiently -- perhaps because of the existence of significant turbulence and shocks throughout the gas. Moreover, \citet{Lisenfeld14} showed that most of these warm H$_{2}$ galaxies have suffered a significant decrease of the molecular gas content and the associated star formation.

Using multi-wavelength data (UV-to-IR), \citet{Bitsakis10, Bitsakis11} examined the dust obscured star formation activity in HCGs. To interpret their results, an evolutionary sequence was proposed, according to the fraction of early-type galaxies each group contained. Groups with more than 25 per cent early-type galaxies were classified as ``dynamically old'', while the remaining groups (hosting mostly spiral and irregular galaxies) as ``dynamically young''. They showed that late-type galaxies in dynamically old groups tend to exhibit redder UV-optical colours, after correcting for dust attenuation. These galaxies also display more than an order of magnitude lower specific star formation rates (SSFRs) and higher stellar masses (M$_{star}$), than the corresponding galaxies found in dynamically young groups as well as galaxies in other environments (such as the field and early-stage interacting pairs). Moreover, \citet{Bitsakis14} examined the cold dust emission in HCGs and showed that late-type galaxies in dynamically old groups display high dust deficiencies (almost an order of magnitude lower dust-to-stellar mass ratios than dynamically young group and isolated field late-type galaxies). 

Despite this progress, since HCGs are a local galaxy sample, they cannot be used to unravel the evolutional history of galaxies in groups. For such an undertaking, a much larger sample covering a wider range of redshifts is necessary. In \citet{Bitsakis15} we presented the largest multi-wavelength compact group sample to-date. Having UV-to-IR data for 7417 compact group galaxies, at 0.01$<$z$<$0.23, this sample is sufficient to study the evolution of galaxy properties over the past 3 Gyr. In that paper, we examined the evolution of the nuclear activity and found that the fraction of group-affiliated galaxies hosting an Active Galactic Nucleus (AGN) increase significantly with time, even though their overall nuclear luminosities are reduced at that period. We also identified a connection between the incidence of AGN activity and the dynamical state of the group, with galaxies in dynamically old groups being more likely to host an AGN at a given stellar mass. In our current study we use the same sample to explore the evolution of star formation as well as the physical mechanisms that might affect it. More specifically, in \S2 we describe the sample selection, data acquisition and the method of calculating the properties of the individual galaxies. In \S3 we present our results, and in \S4 we discuss their implications in the context of galaxy evolution and propose an new evolutionary scheme for groups. Finally, in \S5 we summarize the main conclusions of our study.

Throughout this work, a flat $\Lambda$CDM cosmological model is used, with parameters: H$_{0}$=70 km s$^{-1}$ Mpc$^{-1}$, $\Omega_{m}$=0.30 and $\Omega_{\Lambda}$=0.70.

\section{Sample selection}
Our sample comprises 1770 compact groups, containing 7417 galaxies. The data acquisition and the spectral energy distribution (SED) fitting, were described in detail in \citet{Bitsakis15}. In the following paragraphs, we present only a brief discussion of the selection process, the morphological classification, the SED fitting as well as the analysis and treatment of possible biases. 

The sample originates from the so-called Catalogue-A of \citet{McConnachie09}. These authors initially improved the compact group selection criteria introduced by \citet{Hickson82}, and then applied them to the whole Sloan Digital Sky Survey ({\it SDSS}) Data Release 7 \citep{Abazajian09}, and compiled their Catalogue-A (containing 2217 compact groups). In that manner, the catalogue is more complete that the original compact group sample of \citet{Hickson82}, as it includes all galaxies that satisfy the selection criteria out to a z$\sim$0.25. To address the goals of our current study of examining the evolution of the nuclear and star formation activities in compact groups, a multi-wavelength coverage of the sample is essential. We cross-correlated Catalogue-A  with the ultraviolet (UV) all sky survey of Galaxy Evolution Explorer  \citep[{\it GALEX};][]{Morrissey05}, as well as the near-infrared (NIR) and mid-infrared (MIR) surveys of 2MASS \citep{Skrutskie06} and {\it WISE} \citep{Wright10} respectively. Since the only selection criterion we applied was the availability of UV-to-MIR observations for a group as opposed to an individual galaxy, our sub-sample is also complete as the original Catalogue-A. Due to SDSS fiber collision constraints, we recovered optical spectroscopy only for 4208 galaxies in our sample (57 per cent of the total sample). This constraint could introduce possible biases in our sample, since the lack of optical spectroscopy will be preferentially found in denser groups. However, since the original compact group sample selection by \citet{McConnachie09}, did not rely on spectroscopic data but instead used simulations to fine tune the brightness and compactness criteria in order to establish group membership, this is not an issue. As an additional precaution and after the final selection of the sample, you use spectroscopic information to reject galaxies as group members if they were obvious cases of interlopers (having line-of-sight velocity differences $>$1200 km sec$^{-1}$ from the brightest group member). Our final sample consists of 7417 galaxies, found in 1770 compact groups (CGs), with N$\ge$3 members. As it was shown in \citet{Duplancic13} and \citet{Sohn16}, galaxy triplets and compact groups with N$\ge$4 have similar properties, which are very different from those observed in galaxy pairs and clusters. Finally, for the purposes of this study, galaxies with no available spectroscopy were assigned with the redshift of their brightest neighbor. 

To classify the morphologies of our galaxies, we use the results of \citet{Simard11}. Based on their findings, galaxies with fitted radial parameters of bulge-to-disk ratio (B/T) $\ge$0.35 and image smoothness parameter at half-light radius (S2) $\ge$0.75, were classified as early-type galaxies (hereafter ETGs; typically lenticular and elliptical galaxies), whereas the remaining as late-type galaxies (hereafter LTGs; refers to spiral and irregular galaxies). By applying this classification, we found 3045 LTGs (41 per cent) and 4367 ETGs (59 per cent) in our sample. To examine the validity of this classification, we compared them with independent morphological classifications from the Galaxy Zoo project\footnote{Available at http://www.galaxyzoo.org} \citep{Lintott11} and found that there is a 96 per cent agreement in the classifications of the 3955 galaxies in common. Using our derived morphologies, we also determined the groups' dynamical states, as described in \citet{Bitsakis10}. Groups with more than 25 per cent of ETGs are classified as dynamically old, and the rest as dynamically young. We find that 373 of our groups are dynamically young (21 per cent), and 1397 are dynamically old (79 per cent)

Using the state-of-the art SED code {\sevensize CIGALE} \citep{Noll09}, we fitted the observed UV-to-IR SEDs of our galaxies, and estimated some of their most important physical properties. The code is based on the total energy balance between the energy absorbed in the UV-optical bands and that re-emitted in the IR, and builds SED models which are then compared to the data. The method uses a delayed star formation history (SFH)\footnote{The delayed SFH is defined as: SFR(t) $\propto$ t exp(-t/$\tau_{1}$), where $t$ is the time and $\tau_{1}$ the e-folding time of the star forming activity. A small value of $\tau\sim$1\,Gyr will model a typical early-type galaxy whereas a higher value of 10\,Gyr will provide a more constant SFR with time, typically obtained for late-type galaxies.}, which can provide good estimates\footnote{The goodness of the fit is given by the reduced-$\chi^{2}$. In our study the median $\chi^{2}$ value is 2.37 (with those of the 25th and 75th percentiles being 1.22 and 4.36, respectively).} of the stellar masses (M$_{star}$) and star formation rates (SFR) of the galaxies \citep[for more details see][]{Ciesla15}. This star formation history is a user-provided parameter which can be modified to explore various scenarios. {\sevensize CIGALE} also accounts for the presence of AGN, based on a library of theoretical models from \citet{Fritz06}, and can be used to separate the amount of AGN emission from that due to star formation.

\begin{figure*}
\begin{center}
\includegraphics[scale=0.85]{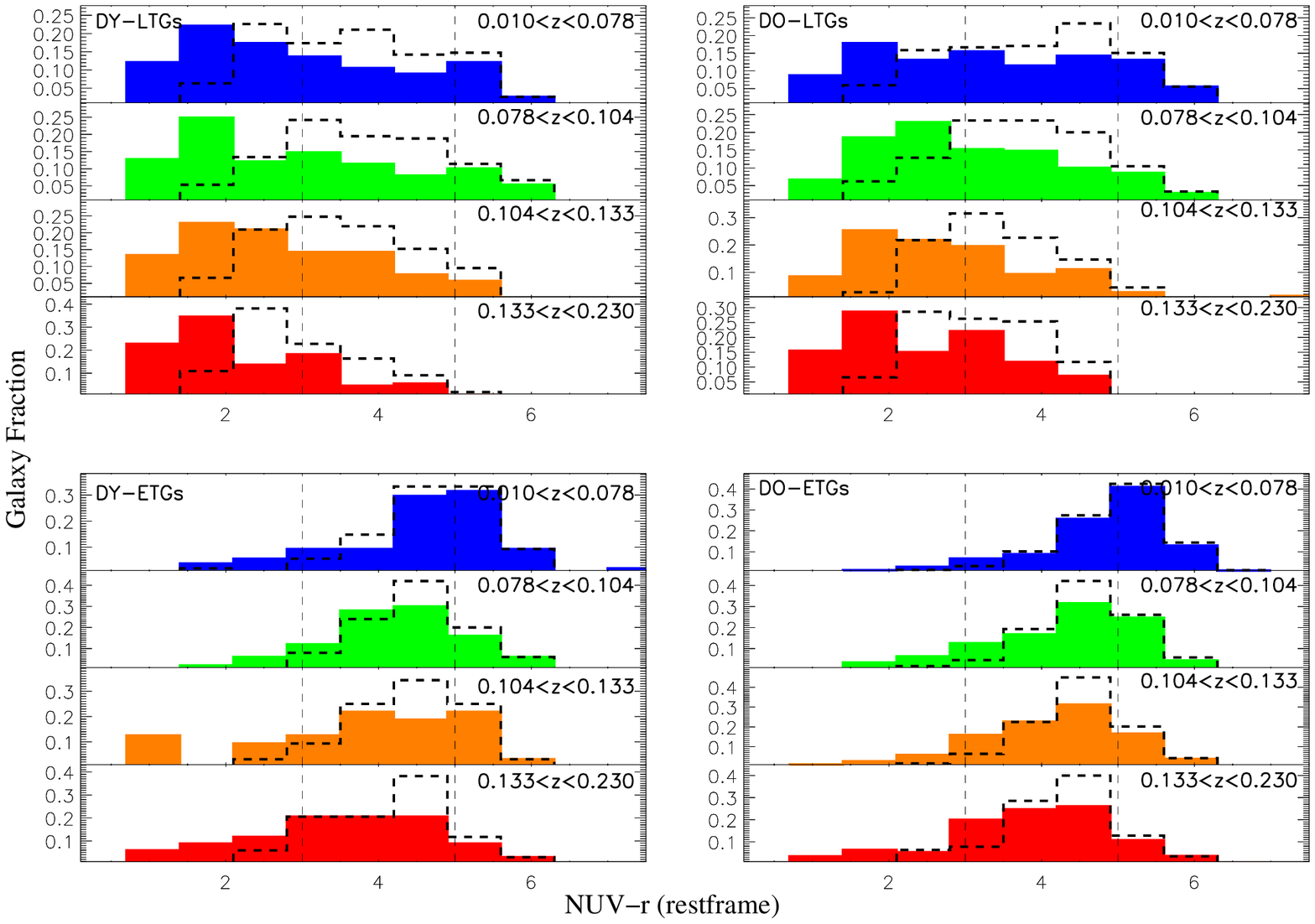}
\caption{Rest-frame [NUV-r] colour distributions of the late-type (LTGs; top panels) and early-type (ETGs; bottom panels) galaxies in dynamically young (DY; left panels) and old (DO; right panels) groups at different redshift ranges. With filled colours we denote the rest-frame [NUV-r] distributions corrected for dust extinction, whereas dashed lines mark the observed ones. The black vertical dashed lines separate the blue cloud ($[NUV-r]<$3), green valley (3$<[NUV-r]<$5), and red sequence ($[NUV-r]>$5) areas of the UV-optical colour-space. To avoid Malmquist bias we only use galaxies in the mass range 10.4$<$ log[M$_{star}$ (M$_{\odot}$)]$<$11.3. (A coloured version of this figure is available on the online journal)}
\label{fig1}
\end{center}
\end{figure*}

To avoid Malmquist bias, when comparing galaxies at different redshifts, we select a mass range that contains galaxies at all redshifts (10.4$<$ log[M$_{star}$ (M$_{\odot}$)]$<$11.3). This covers the typical mass range where compact group galaxies are usually found (estimated to be log(M$_{star}$)$\sim$10.66$\pm$0.74 M$_{\odot}$ in Hickson compact groups; \citealt{Bitsakis11}), and it is used only in the cases of the comparison between galaxies at different redshifts. We also separate our sample into four different redshift bins (Bin1: 0.01$<$z$<$0.078, Bin2: 0.078$<$z$<$0.104, Bin3: 0.104$<$z$<$0.133, and Bin4: 0.133$<$z$<$0.23). The four bins were selected so that they contain an equal number of galaxies ($\sim$1100) after the mass selection criterion was applied, although do not sample the same co-moving volume (for more details see Table 1 of \citealt{Bitsakis15}). 

\section{Results}
\subsection{The evolution of UV-optical colours}

In \citet{Bitsakis11}, we studied the UV-optical colour distribution of 135 galaxies, taken from the original Hickson nearby compact group sample (HCGs). Contrary to what it is observed in the field and in clusters, where a strong bimodality appears between the [NUV-r] colours of star-forming and quiescent galaxies \citep{Wyder07, Haines08}, a large number of the HCG LTGs were located in the intermediate area, the so-called ``green valley'' (having 3$<$[NUV-r]$<$5). A detailed analysis, using SED modeling, ruled out dust extinction as the major mechanism reddening their colours. We also showed that the majority of LTGs located in the green valley, were found in dynamically old groups. We concluded that this behavior could be attributed to past galaxy-galaxy interactions and possible minor merging that may have reduced their star formation activity, by removing gas, as well as, increasing their stellar masses (thus the fraction of red stars).       

\begin{table*}
\begin{minipage}{120mm}
\begin{center}
\caption{Distributions of the observed and the extinction-corrected UV-optical colours of the galaxies in our sample, for different kinds of classified groups.}
\label{tab1}
\begin{tabular}{cccccc}
\hline 
 Redshift bin (Duration$^{a}$ [Gyr])   & Blue cloud (\%) & Green valley (\%) & Red sequence (\%) \\
\hline
\multicolumn{4}{c}{Late-type galaxies in dynamically young groups}\\
                          0.010-0.078 (1.0) & 52$\pm$5 (33$\pm$4)$^{b}$ & 35$\pm$4 (52$\pm$5)$^{b}$ & 13$\pm$2 (15$\pm$2)$^{b}$ \\
\rowcolor{Gray} 0.078-0.104 (0.4) & 50$\pm$6 (26$\pm$5) & 35$\pm$4 (58$\pm$6) & 15$\pm$3 (16$\pm$3) \\
                          0.104-0.133 (0.4) & 57$\pm$7 (38$\pm$6) & 38$\pm$5 (53$\pm$7) & 5$\pm$2 (9$\pm$3) \\
\rowcolor{Gray} 0.133-0.230 (1.0) & 72$\pm$8 (56$\pm$7) & 28$\pm$5 (44$\pm$6) & 0$\pm$0 (0$\pm$0)  \\
\multicolumn{4}{c}{Late-type galaxies in dynamically old groups}\\
                          0.010-0.078 (1.0) & 40$\pm$4 (27$\pm$3) & 41$\pm$4 (54$\pm$4) & 19$\pm$2 (19$\pm$2) \\
\rowcolor{Gray} 0.078-0.104 (0.4) & 49$\pm$5 (28$\pm$3) & 40$\pm$4 (60$\pm$5) & 11$\pm$2 (12$\pm$2) \\
                          0.104-0.133 (0.4) & 56$\pm$5 (38$\pm$4) & 40$\pm$4 (57$\pm$5) & 4$\pm$1 (5$\pm$1) \\
\rowcolor{Gray} 0.133-0.230 (1.0) & 60$\pm$5 (45$\pm$4) & 40$\pm$5 (55$\pm$5) & 0$\pm$0 (0$\pm$0)  \\
\multicolumn{4}{c}{Early-type galaxies in dynamically young groups}\\
                          0.010-0.078 (1.0) & 9$\pm$4  (4$\pm$2) & 54$\pm$9 (57$\pm$10) & 37$\pm$8 (39$\pm$8) \\
\rowcolor{Gray} 0.078-0.104 (0.4) & 8$\pm$4  (2$\pm$2) & 72$\pm$12 (76$\pm$12) & 20$\pm$6 (22$\pm$6) \\
                          0.104-0.133 (0.4) & 22$\pm$8 (9$\pm$5) & 65$\pm$14 (75$\pm$15) & 13$\pm$6 (16$\pm$7) \\
\rowcolor{Gray} 0.133-0.230 (1.0) & 26$\pm$9 (14$\pm$6) & 65$\pm$14 (74$\pm$15) & 9$\pm$5 (12$\pm$5)  \\
\multicolumn{4}{c}{Early-type galaxies in dynamically old groups}\\
                          0.010-0.078 (1.0) & 4$\pm$1 (2$\pm$1) & 48$\pm$2 (47$\pm$3) & 48$\pm$2 (51$\pm$2) \\
\rowcolor{Gray} 0.078-0.104 (0.4) & 10$\pm$1 (3$\pm$1) & 66$\pm$3 (70$\pm$3) & 24$\pm$1 (27$\pm$1) \\
                          0.104-0.133 (0.4) & 10$\pm$1 (4$\pm$1) & 73$\pm$3 (76$\pm$3) & 17$\pm$1 (20$\pm$1) \\
\rowcolor{Gray} 0.133-0.230 (1.0) & 15$\pm$1 (10$\pm$1) & 73$\pm$3 (77$\pm$3) & 12$\pm$1 (13$\pm$1)  \\
\hline
\end{tabular}
\end{center}
{\it Notes:} $^{a}$ Duration of each redshift period, according to $\Lambda$CDM cosmology.\\
$^{b}$In the parentheses appear the fractions of the observed rest-frame colours.
\end{minipage}
\end{table*}%

In Fig.~\ref{fig1}, we present the extinction corrected rest-frame\footnote{The synthetic rest-frame colours are derived from the model-SEDs, using the K-corrections  described in \citet{Chilingarian12}.} [NUV-r] colour distributions of 4355 galaxies in our current sample, within the mass range 10.4$<$ log[M$_{star}$ (M$_{\odot}$)]$<$11.3 (as explained in \S2). We separate them according to their morphology and group dynamical state, as well as the redshift range they are found. The results are given in Table~\ref{tab1}. In the same figure we also include the observed rest-frame distributions (shown with dashed lines). The results suggest a significant colour evolution towards redder [NUV-r] colours, since z$\le$0.23. From Table~\ref{tab1}, we notice that the red-sequence LTGs have been increased by 10-20 per cent and ETGs by $\sim$30 per cent, during the past 3 Gyr.  

Moreover, dynamically old groups contain on average more red galaxies (galaxies found in the green valley and red sequence areas) than the dynamically young groups. This is expected since --by definition-- dynamically old groups contain more ETGs. From Table~\ref{tab1} we can see that ETGs in dynamically old groups are redder than those in dynamically young groups, at all redshifts. What is surprising is that LTGs in dynamically old groups also have redder colours than those found in dynamically young groups. A Kolmogorov-Smirnov analysis reveals that the two LTG distributions are different with P$_{KS}=$ 0.3 per cent\footnote{The Kolmogorov-Smirnov test probability. If less than 1 per cent, the two distributions can be considered as significantly different.}. 

\begin{figure}
\begin{center}
\includegraphics[scale=0.5]{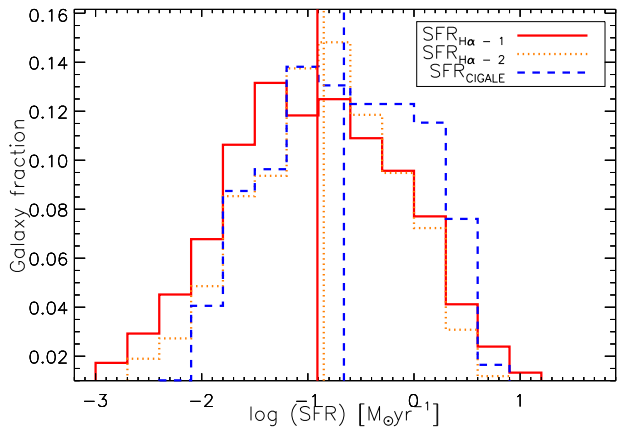}
\caption{Distribution of the star formation rates of the early-type galaxies in our sample, having 10.4$<$ log[M$_{star}$ (M$_{\odot}$)]$<$11.3. With red solid and orange dotted lines we present SFRs estimated using H$\alpha$ fluxes (using the estimators described in \citealt{Kennicutt98} and \citealt{Kennicutt12}, marked with indices 1 and 2 respectively) and with blue dashed lines those from our SED model. The vertical lines indicate the median values of each distribution, respectively. (A coloured version of this figure is available on the online journal)}
\label{fig_TEST}
\end{center}
\end{figure}

Based on the star formation histories of 210 HCG galaxies, \citet{Plauchu12} suggested that most compact groups must have been formed approximately 3 Gyr ago -- also consistent with the hierarchical formation models \citep[e.g][]{Fakhouri10}. Our results, reveal that even though groups at z$>$0.104 do not have any red-sequence LTGs and only a very small fraction ($\sim$10 per cent) of red-sequence ETGs, the evolution of those galaxies towards red colours is rapid. Based on the colours of the different redshift bins we can see that it takes 1 Gyr for the first galaxies to appear in that region of colour-space, with the fastest evolution occurring during the last 1-1.5 Gyr. \citet{Hickson92} showed that a typical timescale over which the gravitational interactions can significantly impact the member galaxies is $\sim$ 200 Myr. This suggests that our galaxies, especially those at lower-z's had already enough time for the multiple interactions and/or minor merging to shape their properties.

Examining the colours of the ETGs we find that in contrast to what is usually observed in clusters \citep[where the majority are found in the red sequence; see][]{Haines08}, less than 30 per cent of our ETGs have such colours. This can be either interpreted as residual star formation, or re-triggering from mergers and/or accumulation of stripped gas. An additional explanation could arise from the UV-upturn phenomenon \citep{Code69, OConnell99}, which is not treated properly by the stellar population synthesis models. According to this, an enhancement of the UV-band emission is observed in some elliptical galaxies, produced by a variety of evolved stars (e.g. old horizontal branch stars) that typically dominate the emission at $\lambda<$2000\AA~\citep{Boselli05} of large elliptical galaxies (with log(M$_{star})>$ 11.4 M$_{\odot}$). This does not seem to be the case here though, because, based on their stellar masses (log(M$_{star})<$ 11.3 M$_{\odot}$; Fig~\ref{fig1}) our ETGs are not giant ellipticals. However, in order to completely rule out emission from evolved stars, we also check the validity of our model-derived SFRs (presented in dashed blue lines in Fig.~\ref{fig_TEST}) by comparing them with those derived using different semi-empirical methods. The comparison is performed with SFRs estimated from H$\alpha$ line fluxes, which are not sensitive to UV-upturn, applying two different methods. The first is described in \citet{Kennicutt98} after correcting H$\alpha$ luminosities from \citet{Dominguez12} (marked as SFR$_{H\alpha - 1}$ in the same figure). The second estimator (marked as SFR$_{H\alpha - 2}$) uses the updated H$\alpha$ SFRs from \citet{Kennicutt12}, applying the extinction corrections discussed in \citet{Garn10}. The small differences among the three estimators (model-derived SFRs are higher by $\sim$0.3dex, independent of the redshift bin), suggest that the UV-upturn is not an important effect in our sample, and therefore the blue [NUV-r] colours are likely due to weak star formation activity. In \S3.3 we describe possible scenarios to conserve and/or re-initiate the star formation in early-type systems. 

\begin{figure*}
\begin{center}
\includegraphics[scale=0.72]{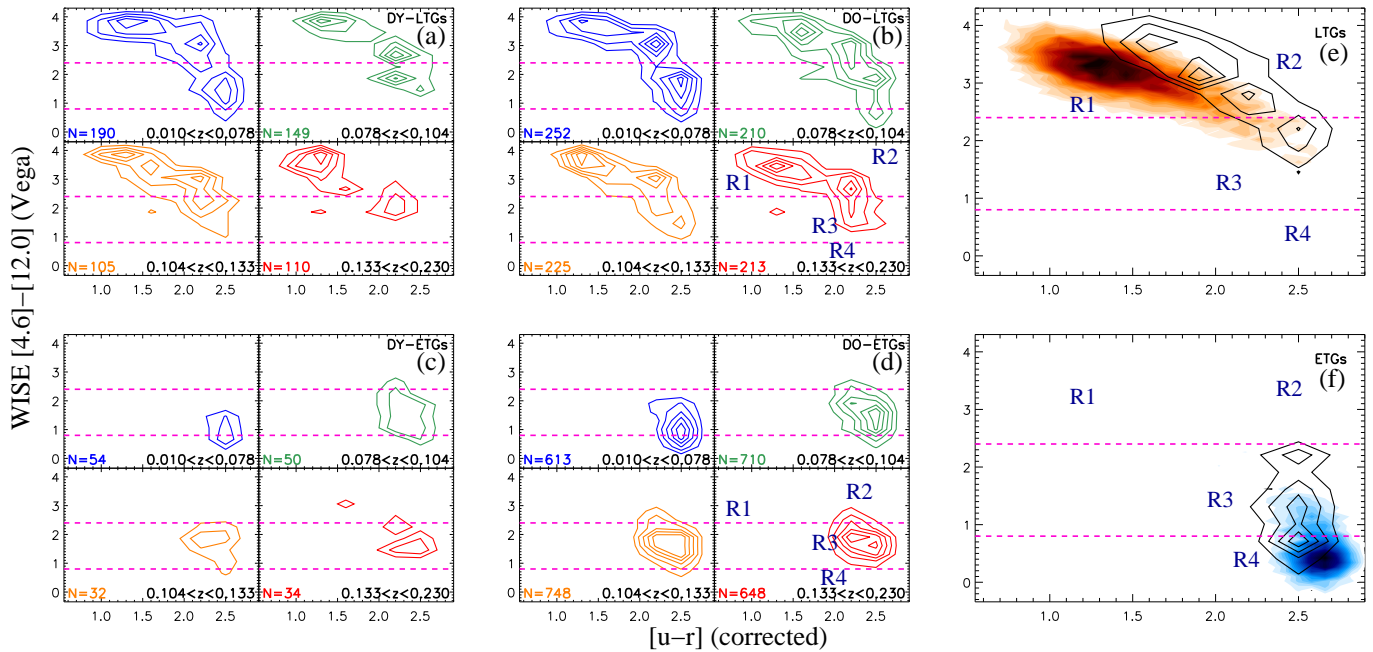}
\caption{The extinction corrected rest-frame $[u-r]$ versus the {\it WISE} $[4.6]-[12.0]\micron$ contours of the galaxies in our sample (panels a-d). The sub-panels indicate the four different redshift bins used in this work, noted in the bottom right corners. At the bottom left corners we mark the number of galaxies in each bin. Contours are in increments of 15 per cent. For comparison we also present the late-type (panel e) and early-type (panel f) galaxies from the Galaxy Zoo \citep[coloured contours; ][]{Schawinski14}, and a cluster galaxy sample \citep[black solid line contours;][]{Lee15}. The dashed magenta lines bracket the infrared transition zone as described in \citet{Alatalo14}. The R1-4 indicate the four regions described in \S3.2. For the purposes of comparison between different redshifts we use galaxies within the mass range of 10.4$<$ log[M$_{star}$ (M$_{\odot}$)]$<$11.3 (A coloured version of this figure is available on the online journal)}
\label{figIRTZ}
\end{center}
\end{figure*}

\subsection{The mid-infrared colours based on WISE}
\citet{Johnson07} examined the distribution of the {\it Spitzer} mid-IR colours of HCG galaxies. They proposed that the ``gap'' in the colours between star forming and quiescent galaxies was only present in compact group samples, and they attributed it to a rapid evolution of the galaxy colours. Although not strictly a gap, as showed by \citet{Bitsakis11}, the properties of the galaxies found in this area were thoroughly studied by \citet{Walker13}, who showed that galaxies in the mid-IR transition region (where that gap was observed) were mainly located in the UV-optical red sequence. These galaxies have been considered a transitional population, based on their star formation rates and molecular gas properties \citep{Lisenfeld14}. 

A couple of other names have been used to describe this colour ``gap'', the ``infrared green valley'' \citep{Cluver13} and the ``infrared transition region'' \citep[IRTZ; ][]{Alatalo14}, when applied to {\it WISE} mid-IR colours. Examining those mid-IR colours of the A2199 supercluster galaxies, \citet{Lee15} proposed an evolutionary scenario where morphological transformation in galaxies occurs while they are found within the mid-IR green valley. 

\citet{Alatalo14} presented the co-evolution of the {\it WISE} mid-IR versus the {\it SDSS} [u-r] colours of several galaxy samples (including star forming, post starburst, and AGN hosting galaxies) and showed that those found in the IRTZ were either classified as Seyferts or low-ionisation emission line regions (LINER), whereas galaxies from the shocked post-starburst galaxy survey \citep[SPOGS;][]{Alatalo16} are located near the star-forming edge, towards red optical colours. They also proposed that galaxies in the IRTZ are mostly late-stage green valley objects transitioning towards red UV-optical colours, as they are shedding their remaining interstellar gas. 

Following the methodology of \citet{Alatalo14}, we present in Fig.~\ref{figIRTZ} the {\it WISE} [4.6]-[12.0]$\micron$ versus the extinction corrected {\it SDSS} [u-r] rest-frame colours of the galaxies in our sample. As in Fig.~\ref{fig1}, we separate galaxies according to their optical morphologies, the dynamical state of their group, and the redshift bin they are found. To compare our observations with a large extragalactic sample that contains galaxies found in various environments (mostly in the field), we present the colour distributions of the 47995 galaxies (65 per cent of which are LTGs) from the Galaxy Zoo sample \citep{Schawinski14}, found at 0.02$<$z$<$0.05 (presented with coloured  contours in panels e and f). We also compare (using black contours) the 1529 late-type (panel e) and early-type (panel f) cluster galaxies of z$\sim$0.03 from \citet{Lee15}, found in the same mass range as our sample. The corresponding fractions of galaxies found in each of the three regions of the mid-IR colour-space are also presented in Table~\ref{tabIRTZ} (for simplicity we do not present the distributions of the [u-r] colours, since they are very similar to those presented in Table~\ref{tab1}). These regions include the active galaxy region, which contains mostly luminous disk galaxies with red colours (with [4.6]-[12.0]$>$2.4) , the IRTZ ( with 0.8 $<$ [4.6]-[12.0]$<$ 2.4), and the passive galaxy region (with [4.6]-[12.0]$<$ 0.8). The latter mostly contains bulge dominates quiescent galaxies. 

\begin{figure*}
\begin{center}
\includegraphics[scale=0.85]{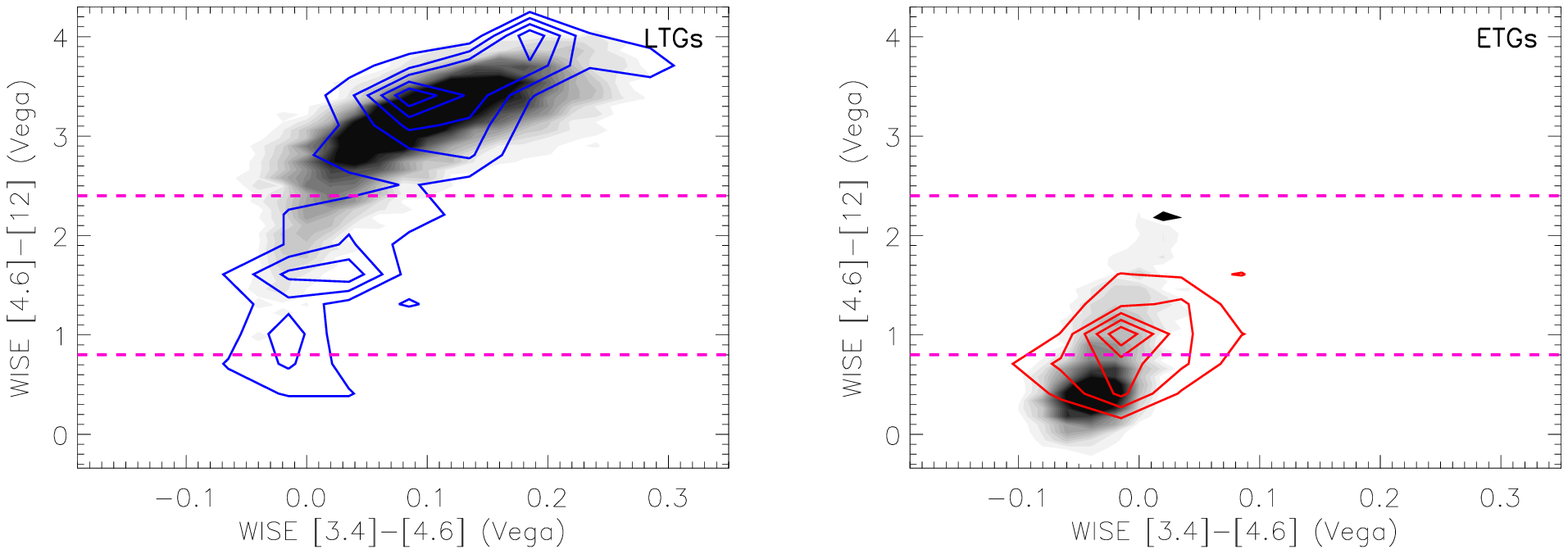}
\caption{The {\it WISE} $[3.4]-[4.6]\micron$ versus the $[4.6]-[12.0]\micron$ contours of the galaxies in our sample at z$<$0.078 (LTGs left panels, ETGs right panel). Contours are in increments of 15 per cent. We also present the colours of the corresponding galaxies in the Galaxy Zoo \citep[black contours][]{Schawinski14}. The dashed magenta lines bracket the infrared transition zone as described in \citet{Alatalo14}. For the purposes of comparison we use galaxies within the mass range of 10.4$<$ log[M$_{star}$ (M$_{\odot}$)]$<$11.3. (A coloured version of this figure is available on the online journal)}
\label{figwalker}
\end{center}
\end{figure*}

\begin{table*}
\begin{minipage}{120mm}
\begin{center}
\caption{Distributions of the WISE [4.6]-[12.0]$\mu$m colours of the galaxies in our sample, as presented in Fig.~\ref{figIRTZ} }
\label{tabIRTZ}
\begin{tabular}{cccccc}
\hline \hline
z-bin  & Active galaxies (\%)$^{a}$ & IRTZ (\%)$^{a}$ & Passive galaxies (\%)$^{a}$\\
\hline
\multicolumn{4}{c}{Late-type galaxies in dynamically young groups}\\
\hline
                          0.010-0.078  & 67$\pm$5 (49$\pm$8)$^{b}$  & 27$\pm$3  & 6$\pm$1 \\
\rowcolor{Gray} 0.078-0.104  & 67$\pm$6 (69$\pm$12) & 28$\pm$4  & 5$\pm$1 \\
                          0.104-0.133  & 69$\pm$8 (36$\pm$8) & 30$\pm$5  & 1$\pm$1 \\
\rowcolor{Gray} 0.133-0.230  & 71$\pm$8 (20$\pm$6) & 27$\pm$4  & 2$\pm$1 \\
\hline
\multicolumn{4}{c}{Late-type galaxies in dynamically old groups}\\
\hline
                          0.010-0.078  & 58$\pm$4 (60$\pm$8) & 35$\pm$3  & 7$\pm$1  \\
\rowcolor{Gray} 0.078-0.104  & 60$\pm$5 (43$\pm$7) & 35$\pm$4  & 5$\pm$1  \\
                          0.104-0.133  & 71$\pm$5 (41$\pm$6) & 28$\pm$3  & 1$\pm$1  \\
\rowcolor{Gray} 0.133-0.230  & 73$\pm$5 (38$\pm$6) & 26$\pm$3  & 1$\pm$1   \\
\hline
\multicolumn{4}{c}{Early-type galaxies in dynamically young groups}\\
\hline
                          0.010-0.078  & 11$\pm$4 (25$\pm$25) & 65$\pm$10  & 24$\pm$6  \\
\rowcolor{Gray} 0.078-0.104  & 14$\pm$5 (100$\pm$66) & 78$\pm$12  & 8$\pm$4  \\
                          0.104-0.133  & 16$\pm$6 (33$\pm$33) & 78$\pm$15  & 6$\pm$4 \\
\rowcolor{Gray} 0.133-0.230  & 35$\pm$10 (83$\pm$37) & 65$\pm$13  & 3$\pm$3   \\
\hline
\multicolumn{4}{c}{Early-type galaxies in dynamically old groups}\\
\hline
                          0.010-0.078  & 8$\pm$1 (100$\pm$21) & 69$\pm$3  & 23$\pm$1  \\
\rowcolor{Gray} 0.078-0.104  & 14$\pm$1 (100$\pm$21) & 78$\pm$3  & 8$\pm$1  \\
                          0.104-0.133  & 16$\pm$1 (100$\pm$20) & 80$\pm$3  & 4$\pm$1  \\
\rowcolor{Gray} 0.133-0.230  & 22$\pm$1 (94$\pm$11) & 76$\pm$3  & 2$\pm$1   \\
\hline
\end{tabular}
\end{center}
{\it Notes:} $^{a}$Active refers to star-forming galaxies with [4.6]-[12.0]$>$2.3 mag, IRTZ to 0.8$<$[4.6]-[12.0]$<$2.3 mag, and Passive to $<$0.8 mag.\\
$^{b}$ The fraction of Active galaxies that are located in the corresponding elbow (Region 2).
\end{minipage}
\end{table*}%

 As it was previously observed in the UV-optical, the strong evolution of galaxy colours is also present in Fig.~\ref{figIRTZ}. Typical LTGs in the field are always located in the upper left portion of the top panels (Region 1; marked with R1 in the same figure). One interpretation of the colour-colour diagram is that as the star formation fades-out, galaxies migrate towards the UV-optical green valley\footnote{ The [u-r] green valley is located between 1.5$<$[u-r]$<$2.5, depending on the absolute V-band magnitude of the galaxies \citep[e.g][]{Haines08}}, yet they still display warm mid-IR colours (Region 2; R2). In the next step, they shift towards the IRTZ (Region 3; R3), which is mostly populated by ETGs (as one can see in the bottom right panel of the same figure). Finally, ETGs move towards the ``red and dead'' galaxy sequence in the bottom right corner of the colour-space (Region 4; R4). Once again we observe that LTGs in dynamically old groups are always few steps ahead in this evolutionary scheme than LTGs in dynamically young groups. 

It is evident that compact group galaxies evolve differently from both comparison samples. Unlike the differences between compact groups and the field, in the percentage of galaxies found in the blue cloud and red sequence, both compact groups and clusters have a large fraction of their galaxies in the green valley-IRTZ region (R3). Contrary to the galaxy populations of Galaxy Zoo, shown in panels e and f of Fig.~\ref{figIRTZ}, the colours of LTGs in compact groups -- irrespective of their dynamical state -- show an ``elbow'' that is observed in the top right of all the sub-figures (R2). One might hypothesize that this is due to enhanced warm dust emission from an AGN. However, {\sevensize CIGALE} is able to remove any dominant AGN component from the {\it WISE} colors \citep[more details of the process in Appendix A and][]{Ciesla15}. AGN-corrected colours are shown as dotted contours in Fig.~\ref{figIRTZAGN}, where we can see that no significant changes occur in the colours of the LTGs. This result was expected since, as showed in \citet{Bitsakis15}, there is a very small fraction ($<$1 per cent) of IR dominant AGN in our sample. Therefore, the elbow in the colour plane can be interpreted as residual star formation. Once this fades-out, elbow galaxies will move towards the IRTZ. Our results are thus in excellent agreement with the findings of \citet{Alatalo15b},  who observed a similar evolution in the corresponding colours of a sample of 14 HCG galaxies. According to their findings, galaxies located in the elbow display star formation suppression (measured using the Schmidt-Kennicutt diagram; \citealt{Kennicutt98}) by a factor of a few. That suppression is increases up to factors of ten (80$\times$ for their most extreme object) as the galaxies shift their colours towards the IRTZ. \citet{Alatalo14} showed that galaxies located in the elbow are mostly classified as Seyferts and post-starburst, whereas those in the IRTZ mostly exhibit LINER emission. An alternative explanation for the creation of the elbow is also discussed by these authors. According to this, it is likely that the [12$\mu$m] band of {\it WISE} will contain a significant contribution of the 11.3$\mu$m polycyclic aromatic hydrocarbon (PAH) feature. It was shown that this emission can be significantly altered by relatively weak UV field \citep[i.e. LINER emission;][]{Sturm06}. However, this band is very wide (covering the [7.1]-[18.1]$\mu$m) and, it may also be affected by the 9.7$\mu$m silicate absorption feature. Thus, in the absence of mid-IR spectra the interpretation on the variation of its strength is rather speculative \citep{Smith07}. 

Examining the distributions of ETGs at higher-z's we can see that the majority is located within the IRTZ, and during the next 2 Gyr their colours will likely evolve towards the lower right corner of the colour-space. The majority of IRTZ galaxies are early-type systems, both in groups and clusters, but the IRTZ still contains a considerable fraction of LTGs ($\sim$10 per cent). It is, thus, possible that these LTGs moved to IRTZ after crossing the star forming and the elbow regions. As a result the morphological transformation of LTGs into ETGs seems to occur within the IRTZ. Our results agree with the conclusions of \citet{Lee15} and \citet{Alatalo15b}, who suggested that LTGs in IRTZ shed their gas content, which has as a result to cease their star formation activity, and to eventually transform them into ETGs. 

Finally, in Fig.~\ref{figwalker}, we present the mid-IR colour distributions of the low-z (z$<$0.078) galaxies of our sample and we compare them with the corresponding distributions of the Galaxy Zoo. As it was expected \citep[see][]{Bitsakis11} the distribution of the LTGs in compact groups is bimodal with the majority located in the mid-IR active galaxy region and a smaller fraction in the IRTZ, just over the ETG galaxy distribution. One can notice the lower concentration of LTGs between the two regions, that is not observed in the LTGs of the field. This effect was initially observed by \citet{Walker10}, who interpreted it as an accelerated evolution of LTGs from star-forming to quiescent. 

\subsection{Star formation activity}

\begin{table*}
\begin{minipage}{120mm}
\begin{center}
\caption{Median star formation properties as a function of redshift, galaxy morphology and dynamical state of the group}
\label{tab2}
\begin{tabular}{ccccc}
\hline \hline
z-bin  & DY-LTGs & DO-LTGs & DY-ETGs & DO-ETGs \\
\hline
\multicolumn{5}{c}{SFR [M$_{\odot}$ yr$^{-1}$]}\\
\hline
                          0.010-0.078  & 1.44 (0.23, 2.81) & 0.94 (0.13, 1.69) & 0.07 (0.01, 0.17) & 0.07 (0.02, 0.13) \\
\rowcolor{Gray} 0.078-0.104  & 2.10 (0.32, 4.01) & 1.47 (0.24, 3.07) & 0.18 (0.03, 0.61) & 0.17 (0.04, 0.36) \\
                          0.104-0.133  & 2.70 (0.74, 5.46) & 3.22 (0.74, 5.00) & 0.49 (0.12, 0.81) & 0.27 (0.10, 0.48) \\
\rowcolor{Gray} 0.133-0.230  & 7.61 (3.13, 11.2) & 4.92 (1.51, 7.74) & 2.18 (0.28, 2.52) & 0.69 (0.24, 1.32) \\
\hline
\multicolumn{5}{c}{SSFR $\times$10$^{-11}$ [yr$^{-1}$]}\\
\hline
                          0.010-0.078  & 2.35 (0.42, 5.05) & 2.29 (0.17, 3.59) & 0.15 (0.03, 0.21) & 0.11 (0.03, 0.17) \\
\rowcolor{Gray} 0.078-0.104  & 3.44 (0.39, 7.32) & 2.33 (0.33, 5.06) & 0.29 (0.06, 1.00) & 0.26 (0.07, 0.57) \\
                          0.104-0.133  & 4.53 (1.17, 8.70) & 4.86 (1.16, 8.01) & 0.60 (0.09, 1.14) & 0.36 (0.12, 0.73) \\
\rowcolor{Gray} 0.133-0.230 & 11.41 (2.86, 20.7) & 7.32 (1.89, 13.8) & 1.63 (0.18, 2.20) & 0.70 (0.23, 1.33) \\
\hline
\end{tabular}
\end{center}
{\it Notes:} In the parentheses appear the 25$^{th}$ and 75$^{th}$ percentiles of each median value, respectively.\\
\end{minipage}
\end{table*}%

\begin{figure*}
\begin{center}
\includegraphics[scale=0.8]{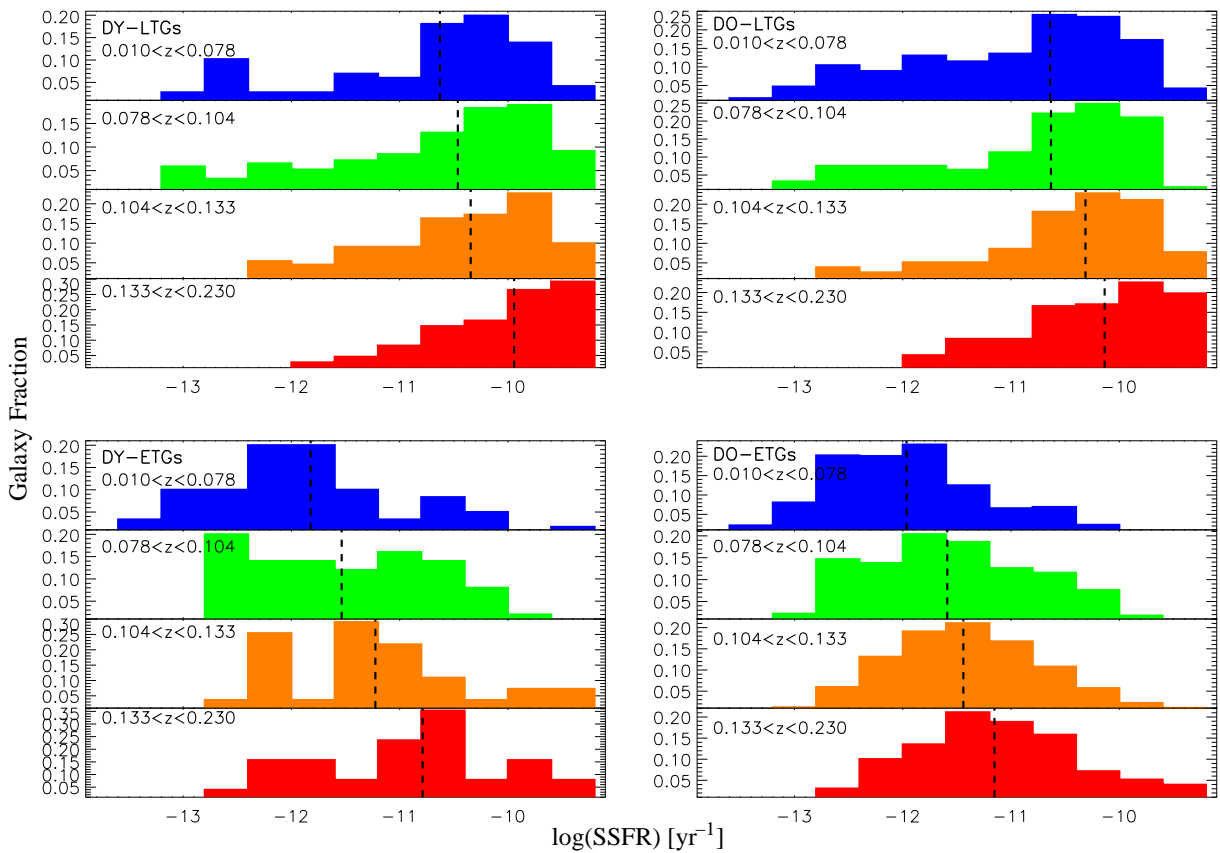}
\caption{Distributions of the specific SFRs (SSFRs) of the LTGs (top panels) and ETGs (bottom panels), found in dynamically young (on the left) and old (on the right) groups, at the different redshift ranges. For the purposes of comparison, we present the distributions of the SSFRs of galaxies in the mass range 10.4$<$ log[M$_{star}$ (M$_{\odot}$)]$<$11.3. (A coloured version of this figure is available on the online journal)}
\label{fig3}
\end{center}
\end{figure*}

In Fig.~\ref{fig3}, we present the distributions of the specific star formation rates (SSFRs) of the galaxies in our sample. Galaxies are separated as in Fig.~\ref{fig1}. In Table~\ref{tab2}, we also present the median values of the SFRs and SSFRs of the sub-samples that are shown in the corresponding panels of Fig.~\ref{fig3}. Consistent with the results of the previous sections, we can observe a significant decrease of the SSFRs throughout time, at all galaxy subsamples. Moreover, galaxies in dynamically old groups appear to have lower median SFRs and SSFRs from those in dynamically young ones, the exception being LTGs of Bin 3.

\begin{figure}
\begin{center}
\includegraphics[scale=0.50]{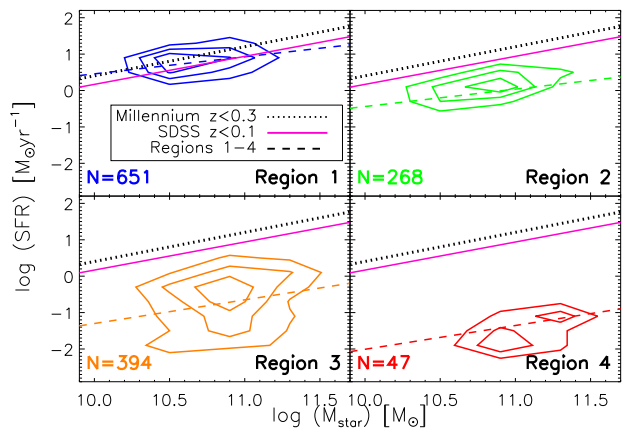}
\caption{ The M$_{star}$--SFR relation of the late-type galaxies in our sample found in the mass range 10.4$<$ log[M$_{star}$ (M$_{\odot}$)]$<$11.3 at all redshifts. The four different panels represent galaxies found in the Regions 1-4, respectively. The magenta solid and the black dotted lines indicate the star forming sequences at z$\le$0.1 and z$<$0.3 \citep[described in ][]{Elbaz07}, whereas the coloured dashed lines are the linear fits of the compact group galaxies in each region, respectively. Contours are intervals of 25 per cent. (A coloured version of this figure is available on the online journal)}
\label{figSFseq}
\end{center}
\end{figure}

\citet{Elbaz07} showed that the star formation-environmental dependence we observe today, where galaxies in dense environments tend to be redder and more quiescent, reverses at z$\sim$1, with the denser environments hosting most of the star forming galaxies. In Fig.~\ref{figSFseq}, we present the stellar mass-SFR relation of the LTGs in our sample, found in the mass range described earlier. The four panels correspond to the regions described in \S3.2. The magenta solid line indicates the star forming main sequence derived for z$\le$0.1 {\it SDSS} galaxies and the black dotted line for z$\le$0.3 galaxies from the {\it Millennium} simulation \citep[both described in][]{Elbaz07}. The dashed lines fit the corresponding compact group data in each panel. Galaxies in  Region 1 follow the star forming sequence, whereas those in Regions 2-4 are moving increasingly away from it, despite the fact that they are still morphologically classified as LTGs -- which naturally follow the main sequence. These results are in agreement with those of \citet{Erfanianfar15}, where they show that, at low-z's, groups contain a large population of red disk-dominated galaxies -- not present in less dense environments -- that are located below the main sequence.  

\begin{figure}
\begin{center}
\includegraphics[scale=0.5]{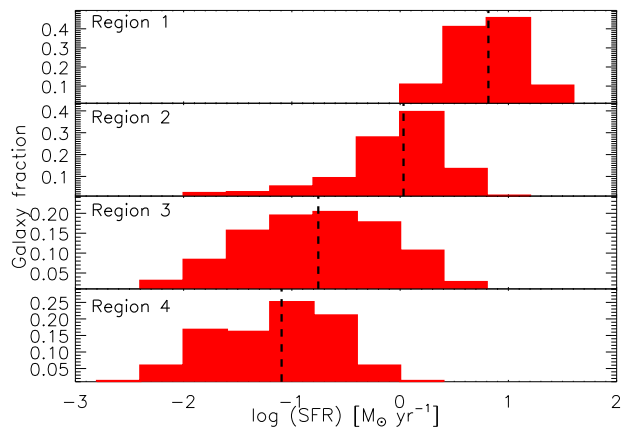}
\caption{Distributions of the SFRs of all the galaxies in our sample within the stellar mass range 10.4$<$ log[M$_{star}$ (M$_{\odot}$)]$<$11.3. The four panels correspond to the different regions of the [u-r] vs WISE colour-space, as presented in \S3.2. (A coloured version of this figure is available on the online journal)}
\label{fig_reg}
\end{center}
\end{figure}

In Fig.~\ref{fig_reg}, we present the distributions of the SFRs of all the galaxies in our sample (both LTG and ETG), separating them to the regions described in \S3.2. We can see that galaxies in Region 1 display significantly higher SFRs than galaxies in Region 2 (the elbow) -- also according to Kolmogorov-Smirnov test, with possibility of P$_{KS}$$<$10$^{-4}$. Despite that galaxies in both Regions 1 and 2 have red WISE colors and are classified as LTGs, the former display SFRs consistent with those of typical LTGs, whereas the SFRs of the latter are already suppressed. In addition, galaxies of Regions 3 and 4 have significantly lower SFRs, with the distribution of the galaxies in Region 3 covering partially both Regions 2 and 4. This result is very interesting given that Region 3 contains both LTGs and ETGs, and may be consistent with the scenario of morphological transformation we described in \S3.2.  

The high SFRs observed in ETGs are not due to excess UV due to the UV-upturn (see \S 3.1), but are rather due to residual star formation activity or re-started star formation due to gas accretion or mergers. Such increase could occur by accretion and/or merging of their companions. In \citet{Bitsakis14} we also found several dusty green-valley ETGs that display similar M$_{H2}$-to-M$_{dust}$ ratios and SSFRs with green valley LTGs. It is worth noticing here we detect them more often at higher z, which may be due to the fact that at earlier epochs, the LTG companions from which they could potentially accrete gas, have higher gas fractions. However, such  accretion may not always stimulate star formation. \citet{Appleton14} presented the case of NGC3226, where the cold accretion from a tidal stream, results in a net inhibition of star formation, by introducing turbulence into the interstellar medium (ISM) of this galaxy. Another, possibility is that our ETGs may have some residual star formation from remaining gas. It is not clear, though, why these galaxies did not consume that gas earlier to form stars. In the next paragraph we will describe mechanisms that inhibit star formation by introducing shocks and turbulence into the ISM of galaxies.

\subsection{The possibility of shocks}
Our results so far, as well as those from the analyses of similar compact group samples \citep[i.e.][]{Alatalo15b}, suggest that the suppression of star formation in LTGs occurs before these galaxies morphologically transform into earlier types. What possible mechanisms can produce such an effect? In addition to gas stripping, which is of a great importance in HCGs \citep[see][]{Verdes01}, several compact group studies argued that shocks and turbulence might be responsible \citep[e.g][]{Cluver13, Lisenfeld14}. Although shocks are transient phenomena \citep[usually lasting 10-100 Myr;][]{Guillard09}, galaxies in groups are expected to experience multiple interactions that may cause an almost constant input of tidal energy in their ISM \citep[][reports typical dynamical times of 200 Myr]{Hickson92}. \citet{Alatalo14} showed that Region 2 is mostly populated by Seyfert and post-starburst galaxies, whereas Region 3 is dominated by galaxies showing LINER or LINER-like emission that may be originated by shocks.

\begin{figure*}
\begin{center}
\includegraphics[scale=0.75]{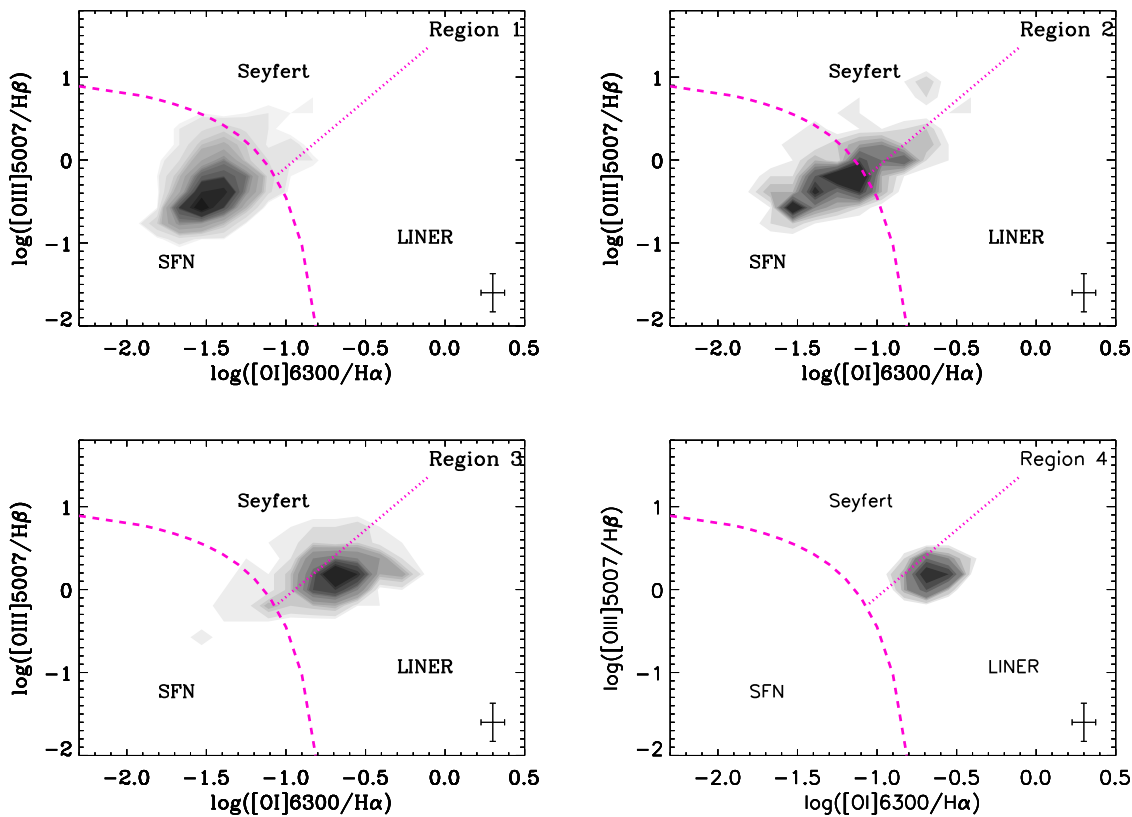}
\caption{Emission line diagnostics from \citet{Kewley06}, using the [O{\sevensize III}]$\lambda$5007/H$\beta$ versus the [O{\sevensize I}]$\lambda$6300/H$\alpha$ line ratios, of the 4208 galaxies in our sample with available optical spectra. The magenta dashed lines separate the star forming nuclei (SFN), from the Seyfert and LINER-like emission ratios. The four panels correspond to the different regions of the [u-r] vs WISE colour-space, as described in \S3.2. Contours are increments of 5 per cent. Typical uncertainties are indicated in the bottom right corner. (A coloured version of this figure is available on the online journal)}
\label{figShocks}
\end{center}
\end{figure*}

However, detecting shocks is a difficult task, especially for a large sample. The kinetic energies of shocks are dissipated through strong radiation fields (in the X-ray and UV bands), resulting in the photoionisation of a large number of species, which affect the global emission line ratios of galaxies. Since no X-ray or UV spectroscopic data are available for our sample, we can only rely on optical spectroscopy. Simulations of \citet{Allen08} showed that shocks can affect the \citet{Kewley06} classification diagrams, making galaxies appear having a LINER-like classification. On the other hand, the emission of shock pre-cursors can imitate that of Seyfert nuclei \citep{Alatalo16}. In Fig.~\ref{figShocks}, we present the optical emission line diagrams, described in \citet{Kewley06}, of the different regions of the [u-r]-WISE colour-space, described in \S3.2. Although, our spectroscopic data  refer to the SDSS single-fiber coverage of the central regions of our galaxies \citep[see Fig. 7 in][]{Bitsakis15}, the significant evolution of the emission line ratios from Regions 1 and 2 towards Regions 3 and 4 is evident. Region 1 mostly contains normal star forming galaxies, in agreement with the results of \S3.3. On the other hand, although Region 2 still contains a large fraction of star forming galaxies (with much lower SFRs, as seen in \S3.3), more than 1/3 of its galaxies are found in the higher excitation region (the threshold between Seyfert and LINER-like emission), suggesting the possibility of on-going shocks in some of these sources, in addition to weak nuclear activity (as described in \citealt{Bitsakis15}). Finally, in addition to AGN, shocks could also explain the behavior of the BPT diagrams of galaxies in both Regions 3 and 4. 

\begin{table*}
\begin{minipage}{120mm}
\begin{center}
\caption{Median star formation, gas properties and inclination of our galaxies as a function of the region of the [u-r]-WISE colour-space they are found}
\label{tab_reg}
\begin{tabular}{ccccc}
\hline \hline
Region  & SSFR$\times$10$^{-11}$ & $\sigma_{Forbidden}$ & inclination & EW(NaD) \\
             &[yr$^{-1}$] & [km sec$^{-1}$] & [$^{o}$] & \AA   \\
\hline
                          Region 1  & 12.21$\pm$0.08 & 75.6$\pm$2.5 & 56$\pm$19 & 1.91$\pm$0.06 \\
\rowcolor{Gray} Region 2  &  2.07$\pm$0.01 & 111.4$\pm$7.3 & 62$\pm$18 & 2.90$\pm$0.19 \\
                          Region 3  &  0.24$\pm$0.01 & 152.7$\pm$8.6 & 46$\pm$19 & 3.34$\pm$0.19 \\
\rowcolor{Gray} Region 4  &  0.07$\pm$0.01 & 159.6$\pm$11.0 & 42$\pm$19 & 3.50$\pm$0.23  \\
\hline
\end{tabular}
\end{center}
\end{minipage}
\end{table*}%

\begin{figure}
\begin{center}
\includegraphics[scale=0.5]{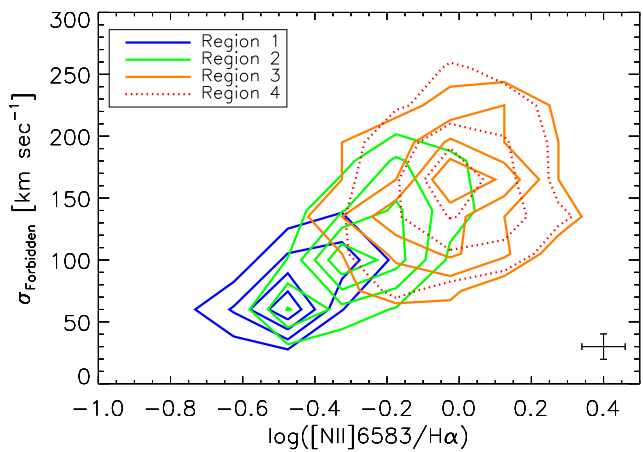}
\caption{Plot of the [NII]$\lambda$6583/H$\alpha$ emission line ratios versus the velocity dispersions of the forbidden lines (i.e. [NII]$\lambda$6583, or [OI]$\lambda$6300), as described in \citet{Brinchmann04}. The blue solid contours correspond to galaxies found in Region 1 of Fig.~\ref{figIRTZ}, green to Region 2, orange to Region 3 and dotted red contours to Region 4. (A coloured version of this figure is available on the online journal)}
\label{figshockcorr}
\end{center}
\end{figure}

If shock excitation is important in our galaxies, then we expect that the effects of shocks will be also reflected in the velocity dispersions of the forbidden higher-excitation lines (i.e. [NII]$\lambda$6583, or [OI]$\lambda$6300). The reason is that shocks will deposit kinetic energy on the gas increasing its turbulence, broadening the emission lines -- especially those of higher excitation \citep[see][]{Rich14,Rich15}. In Fig.~\ref{figshockcorr} we plot the emission line ratios versus the velocity dispersion ($\sigma_{Forbidden}$) of these lines for the galaxies in our sample, again separating them as in \S3.2. We also present their median values in the second column of Table~\ref{tab_reg}. Our results suggest that only galaxies from Regions 3 and 4 (and a very small minority from Region 2) display emission line ratios as well as velocity dispersions that could be consistent with shocks (i.e. $\sigma>$150 km sec$^{-1}$). Nevertheless, the rotation of these galaxies -- especially those with higher inclinations -- could bias the gas velocity dispersions that we measure \citep[enhancing them even by 30 percent; see][]{Bellovary14}. Using the results of the radial profile fitting of our galaxies from \citet{Simard11}, we present in Fig.~\ref{shockstest} and Table~\ref{tab_reg} the inclinations of our galaxies as a function of the velocity dispersion of their gas. From the above it is evident that despite galaxy inclination can be high in many cases, it could not bias the results of Fig.~\ref{figshockcorr}, since it does not prefer any particular region but its distribution is rather random.

In the last column of Table~\ref{tab_reg} we also present the median values of the sodium D (Na{\sevensize D}) equivalent widths (EW(Na{\sevensize D})) of the galaxies in these four regions. Na{\sevensize D} (at 5889 and 5895\AA) is produced in stars -- especially those of spectral types F-G-K-M -- and enrich the ISM via stellar winds, where its depleted on dust grains. Large velocity shocks can destroy those grains, and therefore increase the fraction of gas-phase Na{\sevensize D} in the ISM \citep{Murray07}. Na{\sevensize D} can also probe outflows \citep[see][]{Rupke05, Veilleux05}, and therefore is degenerate \citep{Lehnert11}, from Table~\ref{tab_reg} we can still see that EW(Na{\sevensize D})'s almost double their value, from Region 1 to Region 4. 

AGN feedback could also be an alternative mechanism to explain the suppression of star formation, even if sometimes can also provide positive feedback \citep[see][]{Combes15, Lanz16}. In \citet{Bitsakis15} we showed how the fraction of AGN hosting compact group galaxies increases towards lower redshifts. We also showed that the vast majority are classified as low-luminosity AGN (LLAGN), displaying AGN H$\alpha$ luminosities of $\sim$1-4$\times$10$^{40}$ erg sec$^{-1}$, at z$>$0.104 and decreasing to $<$0.6$\times$10$^{40}$ erg sec$^{-1}$, at z$<$0.078. Although, LLAGN can interact with the molecular content of their hosts suppressing the global star formation \citep[such as the case of NGC1266][]{Alatalo15}, \citet{Alatalo15b} showed that in their sample of star-formation suppressed HCG galaxies the energy introduced to the molecular gas by shocks and turbulence is self-sufficient to stabilize it against gravitational collapse, and therefore no AGN feedback is necessary to explain the significant suppression of star formation observed. 

\section{Discussion: A possible evolutionary sequence for galaxies in groups}
Although the general belief about how galaxies age, whereby LTGs deplete their gas content prior to their quenching and the morphological transformation into ETGs \citep[e.g.][]{Hopkins06}, the results of the current --as well as several recent-- studies suggest that reality is more complicated. We showed that knowing the morphology of a galaxy is not enough to infer its star formation activity and colour. In addition, \citet{Alatalo15b} concluded that even knowing the molecular gas content of a galaxy is not enough to deduce its star formation activity. We have seen that in addition to tidal stripping, galaxies in compact groups might also be subject to shocks and turbulence that can significantly alter their star formation activity. Moreover, we showed that the morphological transformation of compact group late-type galaxies into earlier types occurs not in the UV-optical green valley, as it is currently accepted, but rather in the mid-IR IRTZ. Based on these results, we present a new evolutionary scheme for galaxies in groups. According to this, the evolution of member galaxies is tightly connected with the dynamical evolution of their group. 

Initially, dynamically young groups contain mostly gas-rich late-type galaxies. Later on, the frequent interactions strip significant amounts of gas out of their disks \citep[i.e.][reported HI deficiencies of $>$40 per cent]{Verdes01}, and therefore to reduce their star formation activity. However, the gas in the central region of these galaxies is less affected by the tidal forces, and is not stripped out easily. Therefore, some residual dust obscured star formation may be present. This is supported by the fact that cold dust appears less extended than stars in these galaxies \citep{Bitsakis14}. As a result, during this phase, galaxies will display red UV-optical but red mid-IR colours that can explain the presence of the elbow seen in Fig.~\ref{figIRTZ}. 

As the group evolves dynamically, increasing its velocity dispersion \citep[$\sim$408$\pm$50 km sec$^{-1}$;][]{Bitsakis11} and becoming more compact, galaxies will experience constant harassment, with near-misses and in some cases direct collisions at high speed that can introduce turbulence into their ISM \citep[the case of HCG 57;][]{Alatalo14b}. Moreover, the stripped gas will accumulate in the intra-group space. This will result in the galaxies colliding with this medium. These collisions can induce shocks and turbulence into their ISM, as described in \S3.4. Even though shocks are transient phenomena (10-100 Myr), the frequency of the interactions that trigger them can prevent the ISM of the galaxies from relaxing. This will result in the suppression of star formation in these systems, the fading of which can also lead to the appearance of the weak background AGN activity that $>$40 per cent of these galaxies display \citep{Martinez10, Sohn13, Bitsakis15}. During that period the frequent interactions will also transform the galaxies into earlier type morphologies (as it is seen in the IRTZ; \S3.2). Although, initially that process is slow -- with no significant changes occurring within the first Gyr -- the frequency of galaxy-galaxy and galaxy-debris interactions lead to its acceleration (especially during the last 1-1.5 Gyr; \S3.1, 3.2).

It is also likely that despite their morphology, these early-type galaxies might contain considerable amounts of ISM either from accretion and merging, or residual gas left unable to further form stars because of the shocks. In cases where the cooling processes had enough time to be completed, this will result into the re-ignition of star formation. In those cases, the short-lived burst will change their colours back to ``green''. Eventually, star formation will cease and these galaxies will move to the red galaxy sequence in the bottom right corner of Fig.~\ref{figIRTZ}. 
 
The above evolutionary scheme is consistent with the idea that, although secular evolution is important in transforming galaxies in the field, it is only a minor effect for galaxies in compact groups, where the multiple interactions that a given galaxy will experience during its lifetime will affect more drastically its evolution. 

\section{Conclusions}
In this work we have studied the evolution of star formation and the properties of the galaxies in compact groups, over the past 3 Gyr. Our sample comprises 1770 compact groups (containing 7417 galaxies) found at z=0.01-0.23, and it is the largest multi-wavelength compact group sample to-date. To estimate the properties of the galaxies we have fitted their UV-to-mid-IR spectral energy distributions with {\sevensize CIGALE}. Based on the analysis of these results we conclude that:  

\begin{itemize}

\item The UV-optical and mid-IR colours of our galaxies as well as their star formation activity, evolve differently from galaxies in the field and in clusters. We also observe a significant evolution of these properties with time, accelerating towards lower redshifts. This change appears stronger for galaxies in dynamically old groups. 

\item Late-type galaxies in compact groups found below the star forming main sequence display the highest gas velocity dispersions and excitation lines consistent with shocks. In addition to gas stripping, shocks might also be responsible for inhibiting the star formation in these systems.

\item The morphological transformation of late-type galaxies into earlier types occurs during their transition through the mid-infrared green valley, rather than the UV-optical one.  
 
\item About 70 per cent of the early-type galaxies in compact groups are located within the UV-optical green valley. We rule-out the possibility that the UV-upturn has contaminated our SFR determination, and argue that it is due to residual star formation.  

\item We propose an evolutionary scheme where, in addition to gas stripping, the frequent and multiple interaction along with turbulence and shocks, profoundly shape the morphology and star formation activity of compact group galaxies.

\end{itemize}

\section*{Acknowledgments}
T.B. would like to acknowledge support from the DGAPA-UNAM postdoctoral fellowships. D.D. acknowledges support through grant 107313 from PAIIT-UNAM. V.C. would like to acknowledge partial support from the EU FP7 Grant PIRSES-GA-2012-316788. AZ acknowledges funding from the European Research Council under the European Union's Seventh Framework Programme (FP/2007-2013)/ERC Grant Agreement n. 617001. Y.K acknowledges support from grant DGAPA PAIIPIT IN104215 and CONACYT grant168519. This research has made use of data products from: Galaxy Evolution Explorer (GALEX), and ultraviolet space telescope operated by Caltech/NASA, Infrared Science Archive (IRSA/Caltech), a UCLA/JPL-Caltech/NASA joint project, and Sloan Digital Sky Survey (SDSS). T.B. would also like to thank, P. Bonfini, G. Maravelias, A. Maragkoudakis and A. Steiakaki for the useful discussions/suggestions.

\newpage
\onecolumn
\appendix
\section{Removing the AGN contribution from the photometry of the galaxies}
\begin{figure*}
\begin{center}
\includegraphics[scale=0.85]{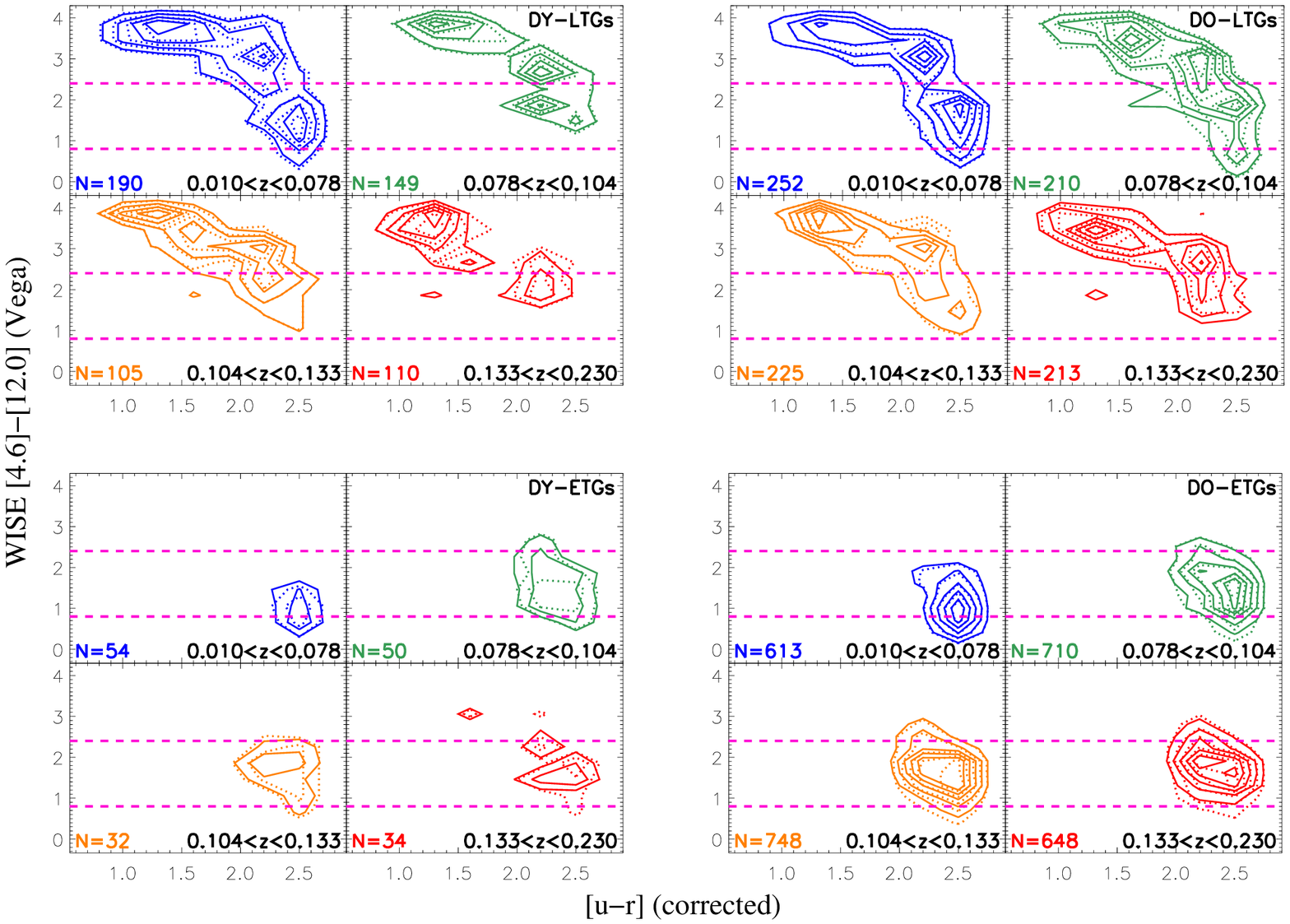}
\caption{The extinction corrected rest-frame $[u-r]$ versus the {\it WISE} $[4.6]-[12.0]\micron$ contours of the galaxies in our sample. With solid contours we present the galaxy colours as estimated by the model, and with dotted contours the same distributions after removing the contribution of the AGN --if present-- from the mid-IR bands. The  panels indicate the four different redshift bins used in this work, noted in the bottom right corners. Contours are in increments of 15 per cent. The dashed magenta lines bracket the infrared transition zone as described in \citet{Alatalo14}. For the purposes of comparison between different redshifts we use galaxies within the mass range of 10.4$<$log(M$_{star}$)$<$11.3 M$_{\odot}$ (A coloured version of this figure is available on the online journal)}
\label{figIRTZAGN}
\end{center}
\end{figure*}

As described in \S2, one of the parameters in the model we are using to fit the observed SEDs of our galaxies, is the contribution from an AGN power-law in the near- and mid-IR bands. This contribution can be specified using AGN Type-1, Type-2 or intermediate type templates, as described in \citet{Ciesla15}, and can be considered only for AGN fractions of f$_{AGN}>$ 10 per cent. Using {\sevensize CIGALE} we remove this contribution -- where present -- from our galaxies, and then with the use of specialized {\sevensize IDL} packages, we estimate their theoretical photometry. The results are shown in Fig.~\ref{figIRTZAGN}. We also perform Kolmogorov-Smirnov analysis between the observed and AGN-removed WISE colours of the galaxies in each panel of this figure (presented in Table~\ref{tabKS}). We can notice that no significant changes occur between the observed (presented with solid contours) and the theoretical colors (in dotted contours), with only exception ETGs in dynamically old groups at z$>$0.133. This result was expected since, as already described in \citet{Bitsakis15}, there are no dust obscured AGN in our sample.   

\begin{table*}
\begin{minipage}{120mm}
\begin{center}
\caption{Kolmogorov-Smirnov tests between the distributions of the observed and AGN-removed WISE colours presented in Fig.~\ref{figIRTZAGN}.}
\label{tabKS}
\begin{tabular}{ccccc}
\hline \hline
z-bin  & DY-LTGs & DO-LTGs & DY-ETGs & DO-ETGs \\
\hline
\multicolumn{5}{c}{Kolmogorov-Smirnov probabilities}\\
\hline
                          0.010-0.078  & 0.99 & 0.99 & 0.99 & 1.00 \\
\rowcolor{Gray} 0.078-0.104  & 0.80 & 0.87 & 1.00 & 0.31 \\
                          0.104-0.133  & 0.91 & 0.18 & 1.00 & 0.02 \\
\rowcolor{Gray} 0.133-0.230  & 0.13 & 0.35 & 0.82 & 0.01 \\
\hline
\end{tabular}
\end{center}
{\it Notes:} The Kolmogorov-Smirnov test probability. If less than 0.01, the two distributions can be considered as significantly different.
\end{minipage}
\end{table*}%

\section{Evaluating the gas velocity dispersions}
\begin{figure*}
\begin{center}
\includegraphics[scale=0.8]{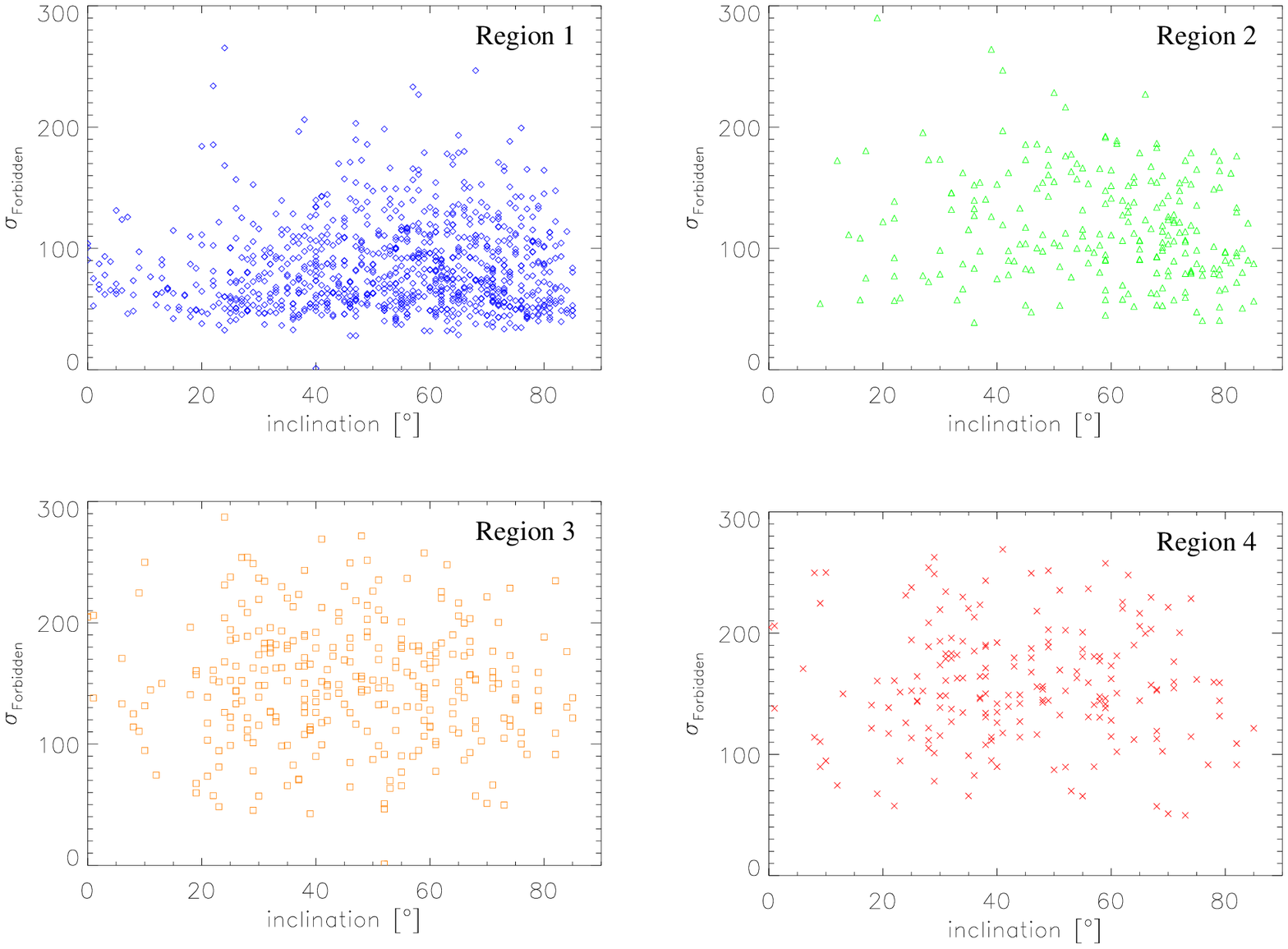}
\caption{ The disk inclination as a function of the velocity dispersion of the forbidden lines for the galaxies in our sample, located in Region 1 (upper left panel), Region 2 (upper right), Region 3 (lower left) and Region 4 (lower right). (A coloured version of this figure is available on the online journal)}
\label{shockstest}
\end{center}
\end{figure*}

To test the validity of the results presented in Fig.~\ref{figshockcorr}, we plot in Fig.~\ref{shockstest} the disk inclination of our galaxies, derived from the radial profile fitting of \citet{Simard11}, as a function of the velocity dispersion of the forbidden lines. The results of this test are also presented in the forth column of Table~\ref{tab_reg}. As one can see there is no correlation between the velocity dispersion of the forbidden lines and the inclination of the galaxies (especially in Regions 3 and 4) that would explain the enhancement of the observer velocities to these regions due to the rotation of these galaxies. Performing a Spearman Rank test between the inclination and the $\sigma_{Forbidden}$ values, we find significance probabilities of 0.33, 0.08, 0.32 and 0.80 respectively (if less than 0.01 there is a significant correlation), suggesting that there is no correlation between the two parameters for any of the four regions we are studying. These results support the conclusions of \S3.4 that the velocity dispersion of the gas increases in Regions 3 and 4 along with the higher excitation line ratios.  

\end{document}